\documentclass[aps,epsf,showpacs,usenatbib]{iopart}
\usepackage{graphics}
\usepackage{graphicx}
\usepackage{epsfig}
\usepackage{times}
\usepackage{subfigure}
\usepackage{psfrag}
\usepackage{amssymb}

\def\reff@jnl#1{{\rm#1\/}}
\def\aj{\reff@jnl{AJ}}                 
\def\araa{\reff@jnl{ARA\&A}}           
\def\apj{\reff@jnl{ApJ}}               
\def\apjl{\reff@jnl{ApJ}}              
\def\apjs{\reff@jnl{ApJS}}             
\def\ao{\reff@jnl{Appl.Optics}}        
\def\apss{\reff@jnl{Ap\&SS}}           
\def\aap{\reff@jnl{A\&A}}              
\def\aapr{\reff@jnl{A\&A~Rev.}}        
\def\aaps{\reff@jnl{A\&AS}}            
\def\azh{\reff@jnl{AZh}}               
\def\baas{\reff@jnl{BAAS}}             
\def\jrasc{\reff@jnl{JRASC}}           
\def\memras{\reff@jnl{MmRAS}}          
\def\mnras{\reff@jnl{MNRAS}}           
\def\pra{\reff@jnl{Phys.Rev.A}}        
\def\prb{\reff@jnl{Phys.Rev.B}}        
\def\prc{\reff@jnl{Phys.Rev.C}}        
\def\prd{\reff@jnl{Phys.Rev.D}}        
\def\prl{\reff@jnl{Phys.Rev.Lett}}     
\def\pasp{\reff@jnl{PASP}}             
\def\pasj{\reff@jnl{PASJ}}             
\def\qjras{\reff@jnl{QJRAS}}           
\def\skytel{\reff@jnl{S\&T}}           
\def\solphys{\reff@jnl{Solar~Phys.}}   
\def\sovast{\reff@jnl{Soviet~Ast.}}    
\def\ssr{\reff@jnl{Space~Sci.Rev.}}    
\def\zap{\reff@jnl{ZAp}}               
\def\nat{\reff@jnl{Nature}}            

\def\cqg{\reff@jnl{Class. Quantum Grav.}} 

\begin{document}
\input epsf.tex

\title[Cosmic Swarms]{Cosmic Swarms: A search for Supermassive Black Holes in the LISA data stream with a Hybrid Evolutionary Algorithm}
\author{Jonathan R. Gair$^{1}$ and Edward K. Porter$^{2,3}$}
\address{$^{1}$ Institute of Astronomy, Madingley Road, CB3 0HA Cambridge, UK} 
\address{$^{2}$ APC, UMR 7164, Universit\'e Paris 7 Denis Diderot,\\ 10, rue Alice Domon et L\'{e}onie Duquet, 75205 Paris Cedex 13, France}
\address{$^{3}$ Albert Einstein Institute, Am M\"uhlenberg 1, D-14476, Golm, Germany}

\vspace{1cm}
\begin{abstract}
We describe a hybrid evolutionary algorithm that can simultaneously search for multiple supermassive black hole binary (SMBHB) inspirals in LISA data.  The algorithm mixes evolutionary computation, Metropolis-Hastings methods and Nested Sampling. The inspiral of SMBHBs presents an interesting problem for gravitational wave data analysis since, due to the LISA response function, the sources have a bi-modal sky solution.  We show here that it is possible not
only to detect multiple SMBHBs in the data stream, but also to investigate simultaneously all the various modes of the global solution.  In all cases, the algorithm returns parameter determinations within $5\sigma$ (as estimated from the Fisher Matrix) of the true answer, for both the actual and antipodal sky solutions.
\end{abstract}

\maketitle

\section{Introduction}
The coalescences of supermassive black holes (SMBHs) in binaries are expected to be a major source of gravitational waves (GWs) for the Laser Interferometer Space Antenna (LISA)~\cite{lisa}.  Due to the huge amount of energy released in gravitational radiation during the final inspiral and merger, these SMBH binaries (SMBHBs) will be detected by LISA at cosmological distances with signal to noise ratios (SNRs) of many hundreds to thousands.  The brightness of these sources should allow us to conduct high precision cosmology and cosmography, at a level of accuracy that is unprecedented in modern astronomy.  We will be able to detect SMBHBs out to a redshift of $z\sim10$~\cite{hugheslang, cornishporter1} with parameter errors of a few percent, and may also be able to use these observations to test the predictions of General Relativity, such as the uniqueness of a Kerr black hole as the endstate of gravitational collapse.

There is a significant effort being made at present in both parameter estimation (see for example~\cite{LISApet}) and the development of search algorithms for these sources~\cite{mldc}.  To date, the most successful search methods have been based on a variant of a Markov Chain Monte Carlo (MCMC).  This is a stochastic search method which is highly efficient in searching through high dimensional parameter spaces.  The most commonly employed MCMC variant, referred to as Metropolis-Hastings Monte Carlo, has been successfully applied to searches for non-spinning SMBHBs in both controlled~\cite{cornishporter1, cornishporter2} studies, and in blind studies in the context of the Mock LISA Data Challenges~\cite{cornishporter3, mldc}.  The algorithm uses a number of directed proposal distributions to accelerate the convergence of the search.  However, this method is limited in that it only looks for one source at any one time.  In addition, the nature of the detector response is such that low-frequency SMBHBs have bi-modal solutions for the sky location.  With the MHMC algorithm, one would need to run multiple chains in order to explore both modes of the solution.

With this in mind, we have investigated a new method based on a mixture of Evolutionary Algorithms, Nested Sampling~\cite{Skilling04,pelleg,feroz08,multinest} and Metropolis-Hastings methods~\cite{metrop53,hast70}.  Evolutionary algorithms (EAs) are becoming a popular method for searching high dimensional parameter spaces.  The idea is to use aspects from evolution, such as birth, death, fitness etc. to evolve a set of organisms within the environment of the parameter space.  Most EAs work along the principle of generating a random population of organisms, and letting them evolve according to particular rules.  At predetermined points during the evolution, the fitness of each individual organism is tested.  The fitter organisms are allowed to survive, while the most unfit members are killed off.  A particular type of EA, the genetic algorithm, has already been used in gravitational wave astronomy in the search for galactic binaries~\cite{ccr}.

The Nested Sampling algorithm was developed as a tool for evaluating the Bayesian evidence. It uses a number of live points to climb through nested contours of increasing likelihood.  The algorithm works on the principle that the colony of live points is constantly moving to areas of higher likelihood --- at each step a point is found of higher likelihood than the minimum in the set and is used to replace the lowest likelihood point of the cluster.  The primary difficulty of Nested Sampling is to efficiently sample points of higher likelihood, but techniques have been developed that can achieve this~\cite{feroz08,multinest}. Nested Sampling techniques have previously been applied to gravitational wave data analysis, in the context of model selection for ground-based gravitational wave detectors~\cite{Veitch:2008wd}, but to our knowledge this work is the first application to LISA.

The main advantage of both EAs and Nested Sampling is that they use a number of organisms, whereas the standard MCMC employs just one chain at a time.  As the population of organisms can fracture and recombine into clusters, the organisms can move through the parameter space looking for multiple modes of a solution.  This, for example, allows us to circumvent the problems arising from the fact that all SMBHBs produce a bi-modal solution in the sky position.

The detection problem for non-spinning SMBH binaries has already been solved, in the sense that several groups have been able to successfully detect and recover parameters for such systems in the Mock LISA Data Challenges. While solving for multiple such binaries simultaneously will be useful, it is not essential to the success of LISA. However, there are other LISA sources, such as extreme-mass-ratio inspirals, for which it has proven much more difficult to accurately recover parameters~\cite{EtfAG,neilemri,BBGPa,BBGPb}. This has been in part because of the large number of secondary modes of the solution that exist in the likelihood surface. Techniques which can simultaneously find multiple modes may be essential for successful identification of these types of signal. We have chosen to illustrate the ideas of evolutionary algorithms in this paper with an application to the non-spinning SMBH problem, since the latter is well understood and hence facilitates direct comparison of these methods to other available techniques.

In this paper, we will discuss some of the general principles of evolutionary algorithms, and a particular implementation of a hybrid algorithm that combines the various elements outlined above. The paper is organised as follows. In Section~\ref{sec:waveform} we describe the gravitational waveform model for non-spinning SMBHBs by defining the describing parameter set.  We also outline the parameters of our test sources and the priors we impose on the parameter space.  In Section~\ref{sec:lisada} we define some of the important quantities used in LISA data analysis, such as the Fisher information matrix, the signal-to-noise ratio and the detector noise model that we employ.  In Section~\ref{sec:mhmc} we review the Metropolis-Hastings Monte Carlo method and outline the advantageous features of this algorithm including simulated and thermostated annealing.  In Section~\ref{sec:gbmhmc} we discuss some of the techniques that allow us to go beyond the Metropolis-Hastings algorithm.  This includes a description of Nested Sampling, Metropolis-Hastings Nested Sampling, Evolutionary algorithms and the clustering methods used to partition the live point set. In Section~\ref{sec:hea} we describe our hybrid evolutionary algorithm, before presenting in Section~\ref{sec:heares} the results obtained from using the algorithm to search data sets containing multiple SMBHBs. We finish in Section~\ref{sec:discuss} with a discussion of possible future applications of these techniques.

\section{The gravitational waveform model}\label{sec:waveform}
The waveform for a binary system composed of two Schwarzschild black holes is described by the nine parameter set $\vec{x}=\{\ln(M_{c}),\ln(\mu),\theta, \phi, \ln(t_{c}), \iota, \varphi_{c}, \ln(D_{L}), \psi\}$, where $M_{c}$ is the chirp mass, $\mu$ is the reduced-mass, $(\theta,\phi)$ are the sky location of the source, $t_{c}$ is the time-to-coalescence, $\iota$ is the inclination of the orbit of the binary to the line of sight from the observer, $\varphi_{c}$ is the phase of the GW at coalescence, $D_{L}$ is the luminosity distance and $\psi$ is the polarization of the GW.  The four parameter subset, $\{D_{L}, \iota, \varphi_{c}, \psi\}$, are extrinsic parameters (i.e. parameters that only affect how the gravitational waveform projects onto a detector response), while all the rest are intrinsic ( i.e. parameters that describe the GW phasing at the detector).  We point out that $(\theta,\phi)$, which are normally classed as extrinsic parameters for ground based GW observations, are intrinsic parameters due to the fact that they determine the beam pattern functions and Doppler motion, which are time-dependent due to the motion of LISA over the course of an inspiral observation.
 
In this study, we use the restricted post-Newtonian approximation, i.e. we keep only the dominant Newtonian term in the GW amplitude. Using the low frequency approximation~\cite{cutler}, the detector response is given by
\begin{equation}
h(t) = h_{+}(\xi(t))F^{+}+h_{\times}(\xi(t))F^{\times}, 
\end{equation}
where the phase shifted time parameter is 
\begin{equation}
\xi(t) = t - R_{\oplus}\sin\theta\cos\left(\alpha(t) - \phi\right),
\end{equation}
$R_{\oplus} = 1 AU \approx$ 500 secs is the radial distance to the detector guiding center, $\alpha(t)=2\pi f_{m}t + \kappa$, $f_{m}=1/year$ is the LISA modulation frequency,  $\kappa$ gives the initial ecliptic longitude of the guiding center and  $F^{+,\times}(t)$ are the beam pattern functions~\cite{cornishrubbo}.  The GW polarizations are defined by
\begin{eqnarray}
h_{+}& =& \frac{2Gm\eta}{c^{2}D_{L}}\left(1+\cos^{2}\iota\right)x\cos(\Phi),\\ \nonumber \\
h_{\times} &= &-\frac{4Gm\eta}{c^{2}D_{L}}\cos\iota\,x\sin(\Phi),
\end{eqnarray}
where $m=m_{1}+m_{2}$ is the total mass of the binary, $\eta = m_{1}m_{2}/m^{2}$ is the reduced mass ratio, $G$ is Newton's constant and $c$ is the speed of light.  The invariant PN velocity parameter is $x = \left(Gm\omega / c^{3}\right)^{2/3}$. Here, $\omega$ is the orbital frequency for a circular orbit, which is formally defined as $\omega=d\Phi_{orb}/dt$ and $\Phi =\varphi_{c}-\varphi(t) = 2\Phi_{orb}$ is the gravitational wave phase. We take these at 2PN order, using the expressions~\cite{biww}
\begin{eqnarray}
\omega(t)&=&\frac{c^{3}}{8Gm}\left[\Theta^{-3/8}+\left(\frac{743}{2688}+\frac{11}{32}\eta\right)\Theta^{-5/8}-\frac{3\pi}{10}\Theta^{-3/4}\right.\nonumber\\ &+&\left.\left(\frac{1855099}{14450688}+\frac{56975}{258048}\eta+\frac{371}{2048}\eta^{2}\right)\Theta^{-7/8}\right]\label{eqn:freq}\\
\Phi(t) &=& \varphi_{c}-\frac{2}{\eta}\left[\Theta^{5/8}+\left(\frac{3715}{8064}+\frac{55}{96}\eta\right)\Theta^{3/8}-\frac{3\pi}{4}\Theta^{1/4}\right.\nonumber\\ &+&\left.\left(\frac{9275495}{14450688}+\frac{284875}{258048}\eta+\frac{1855}{2048}\eta^{2}\right)\Theta^{1/8}\right],\label{eqn:phase}
\end{eqnarray}
where
\begin{equation}
\Theta(t;t_{c}) = \frac{c^{3}\eta}{5Gm}\left(t_{c}-t\right).
\end{equation}
The chirp mass, $M_{c}$, and reduced mass, $\mu$ are given in terms of $m$ and $\eta$ by $M_{c}=m\eta^{3/5}$ and $\mu = m\eta$.

\paragraph{Sources Being Investigated}
In Table~\ref{tab:params} we list the parameter values for the two MBHBs that we use in this analysis.  We decided to simultaneously search for one coalescing and one non-coalescing source.   The individual masses in the table are the source frame masses.  We chose very wide priors that encompassed both sources at the same time.  These priors are presented in Table~\ref{tab:priors}.  The relation between luminosity distance and redshift is calculated using the usual relation
\begin{equation}
D_{L} = \frac{c(1+z)}{H_{0}}\int_{0}^{z}\,dz'\left[\Omega_{R}\left(1+z'\right)^{4}+\Omega_{M}\left(1+z'\right)^{3}+\Omega_{\Lambda}\right]^{-1/2}.  
\end{equation}
where we use the concordance WMAP values of $(\Omega_{R}, \Omega_{M}, \Omega_{\Lambda}) = (4.9\times10^{-5}, 0.27, 0.73)$ and $H_{0}$=71 km/s/Mpc~\cite{wmap}.  The luminosity distances for the two sources are then 6.634Gpc and 5.024Gpc respectively.  The redshifted chirp mass and mass ratio are $(4.9289, 1.8182)\times10^6\,M_{\odot}$ for source 1, and $(2.997, 1.44)\times10^6\,M_{\odot}$ for source 2.  The last stable orbit frequencies of the sources are $1.586\times10^{-4}$Hz and $8.9\times10^{-5}$Hz respectively.  The signal to noise ratios (SNR), which we will define in the next section, are 200 for the coalescing source and 131 for the non-coalescing source.

\begin{table}
\caption{\label{tab:params}Parameter values for the two SMBHBs considered in this study.  Note that we quote the individual masses in the rest-frame, not the redshifted masses.} 

\begin{indented}
\lineup
\item[]\begin{tabular}{@{}*{10}{l}}
\br                              
$$&$m_1/M_{\odot}$&$m_2/M_{\odot}$&$tc/yrs$&$\theta$&$\phi$&$z$&$\iota$&$\psi$&$\varphi_c$\cr 
\mr
1&$1\times10^7$ &$1\times10^6$&$0.90$&$0.6283$ &4.7124&1.0&1.1120&1.2330&2.2220\cr
2&$4\times10^6$&$1\times10^6$&$1.02$&$2.1206$&3.9429 & 0.8& 0.6565&2.6354&4.6532\cr 
\br
\end{tabular}
\end{indented}
\end{table}

\begin{table}
\caption{\label{tab:priors}Parameter priors used in the search.  The prior for the total mass is given for the redshifted total mass.} 

\begin{indented}
\lineup
\item[]\begin{tabular}{@{}*{4}{l}}
\br                              
$$&$m_1/m_2$&$m(z)/M_{\odot}$&$tc/yrs$\cr 
\mr
$min$&$15.0$ &$5.0\times10^6$&$0.85$\cr
$max$&$1.0$&$2.5\times10^7$&$1.10$\cr 
\br
\end{tabular}
\end{indented}
\end{table}

\section{LISA data analysis}\label{sec:lisada}
Using a geometric model of signal analysis~\cite{helst, owen}, the waveforms can be thought of as inhabiting a vector space with the natural scalar product
\begin{equation}\label{eqn:scalarprod}
\left<h\left|s\right.\right> =2\int_{0}^{\infty}\frac{df}{S_{n}(f)}\,\left[ \tilde{h}(f)\tilde{s}^{*}(f) +  \tilde{h}^{*}(f)\tilde{s}(f) \right].
\label{eq:scalarprod}
\end{equation}
and vector norm $|h| = \left<h\left|h\right.\right>^{1/2}$, where
\begin{equation}
\tilde{h}(f) = \int_{-\infty}^{\infty}\, dt\, h(t)e^{2\pi\imath ft}
\end{equation}
is the Fourier transform of the time domain waveform $h(t)$.  The quantity $S_{n}(f)$ is the one-sided noise spectral density of the detector.  The total noise power spectral density is in two parts : instrumental noise and galactic confusion noise.  The one-sided noise spectral density for the LISA instrument noise is~\cite{cornish}
\begin{eqnarray}
S_{n}^{\rm inst}(f) &= &\frac{1}{4L^{2}}\left[ 2S_{n}^{\rm pos}(f) \left(2+\left(\frac{f}{f_{*}}^2\right)\right)\right. \nonumber\\
&+ & \left.  8 S_{n}^{\rm acc}(f) \left(1+\cos^2\left(\frac{f}{f_{*}}\right)\right) \left( \frac{1}{(2\pi f)^{4}} + \frac{(2\pi 10^{-4})^{2}}{(2\pi f)^{6}}   \right)   \right] . 
\end{eqnarray}
where $L=5\times10^{6}$ km is the arm-length for LISA,  $S_{n}^{\rm pos}(f) = 4\times10^{-22}\,m^{2}/$Hz and $S_{n}^{\rm acc}(f) = 9\times10^{-30}\,m^{2}/s^{4}/$Hz are the position and acceleration noise respectively.  The quantity $f_{*}=1/(2\pi L)$ is the mean transfer frequency for the LISA arm.  Note that we use a pessimistic noise curve which rises steeply at frequencies lower than $10^{-4}$ Hz.

To model the foreground confusion noise from  galactic white-dwarf binaries, we use the following confusion noise estimate derived from the galaxy model of Nelemans, Yungelson and Portegies-Zwart~\cite{NYZ, TRC}
\begin{equation}
S_{n}^{\rm conf}(f) = \left\{ \begin{array}{ll} 10^{-44.62}f^{-2.3} & 10^{-4} < f\leq 10^{-3} \\ \\ 10^{-50.92}f^{-4.4} & 10^{-3} < f\leq 10^{-2.7}\\ \\ 10^{-62.8}f^{-8.8} &  10^{-2.7} < f\leq 10^{-2.4}\\ \\ 10^{-89.68}f^{-20} &  10^{-2.4} < f\leq 10^{-2}  \end{array}\right.,
\end{equation}
where the confusion noise has units of $m^{2}$Hz$^{-1}$.  In the search, we weight the scalar product, Eq.~(\ref{eq:scalarprod}), with the total power spectral density
\begin{equation}
S_{n}(f) = S_{n}^{\rm ins}(f)+ S_{n}^{\rm conf}(f).
\end{equation}

At low frequency, the three arms of the LISA interferometer can be regarded as being two separate $90^{o}$ interferometers, with the signal in each of the detectors given by
\begin{equation}
s_{i}(t) = h_{i}(t)+n_{i}(t),
\end{equation}
where $i=I, II$ label the detectors. We assume that the noise $n_{i}(t)$ is stationary, Gaussian and uncorrelated in each detector, and characterized by the noise spectral density $S_{n}(f)$.  Using the scalar product, Eq.~(\ref{eqn:scalarprod}), the optimal matched filtering SNR in each detector is
\begin{equation}
\rho_{i} = \frac{\left<h\left|s_{i}\right.\right>}{\sqrt{\left<h\left|h\right.\right>}}.
\end{equation}
Given some signal $s(t)$, the likelihood that the true parameter values are given by a parameter vector $\vec{x}$ is
\begin{equation}\label{eqn:likelihood}
{\mathcal L}\left(\vec{x}\right) = C\,e^{-\left<s-h\left(\vec{x}\right)|s-h\left(\vec{x}\right)\right>/2},  
\end{equation}
where $C$ is a normalization constant.  The maximum likelihood corresponds to the parameter set that minimizes the exponent in the above equation.  As most SMBHBs in LISA will have high SNR, the errors in the parameter estimation at the maximum will have a Gaussian probability distribution given by
\begin{equation}
p\left(\Delta\vec{x}\right)=\sqrt{\frac{\Gamma}{2\pi}}e^{-\frac{1}{2}\Gamma_{\mu\nu}\Delta x^{\mu}\Delta x^{\nu}},
\end{equation}
where $\Gamma_{\mu\nu}$ is the Fisher information matrix (FIM)
\begin{equation}
\Gamma_{\mu\nu} = \left<\frac{\partial h}{\partial x^{\mu}}\left|\frac{\partial h}{\partial x^{\nu}}\right.\right>,
\end{equation}
and $\Gamma = {\rm det}\left(\Gamma_{\mu\nu}\right)$.  When we use both LISA detectors the total SNR is given by
\begin{equation}
\rho = \sqrt{\rho_{I}^2 + \rho_{II}^2},
\end{equation}
while the total FIM is given by
\begin{equation}
\Gamma_{\mu\nu} = \Gamma_{\mu\nu}^{I} + \Gamma_{\mu\nu}^{II}.
\end{equation}
The inverse of the Fisher matrix gives the variance-covariance matrix
\begin{equation}
C^{\mu\nu} = \Gamma_{\mu\nu}^{-1}.
\end{equation}
The diagonal elements of $C^{\mu\nu}$ is a $1\sigma^{2}$ estimate of the error in the recovered parameter
\begin{equation}
\Delta x^{\mu} = \sqrt{C^{\mu\mu}},
\end{equation}
while the off-diagonal elements can be used to obtain the correlations between the various parameters
\begin{equation}
c^{\mu\nu} = \frac{C^{\mu\nu}}{\sqrt{C^{\mu\mu}C^{\nu\nu}}}\,\,\,\,\,\,\,\,\,\,\,\,-1\leq c^{\mu\nu}\leq 1.
\end{equation}

As described earlier, it is possible to divide the parameter space for SMBHB inspirals into intrinsic parameters, $M_c,\mu,\theta,\phi,t_c$, and extrinsic parameters, $D_L, \iota,\phi_c,\psi$. It is possible to maximize the likelihood over the extrinsic parameters analytically using a generalisation of the F-statistic~\cite{JKS}. The details of this maximization for SMBH binary systems are given in~\cite{cornishporter1} so we do not repeat it here. However, we do make use of the F-statistic in our search in order to reduce the dimensionality of the parameter space that we must search.

\section{Metropolis-Hastings Monte Carlo}\label{sec:mhmc} 
In previous works~\cite{cornishporter1,cornishporter2,cornishporter3} a version of Markov Chain Monte Carlo (MCMC), called Metropolis-Hastings Monte Carlo (MHMC), has been used to successfully search for non-spinning SMBHBs.  This variant employs a number of different techniques during the search phase which contravene the Markovian property normally required by standard MCMCs.  Starting with the signal $s(t)=h(t)+n(t)$, we choose a starting point randomly in the parameter space, $\vec{x}$, with template $h(t;\vec{x})$.  We then draw from a proposal distribution, $q(\vec{x}|\vec{y})$, and propose a jump to another point in the space $\vec{y}$.  In order to evaluate whether or not we move, we calculate the Metropolis-Hastings ratio
\begin{equation}
H = \frac{\pi(\vec{y})p(s|\vec{y})q(\vec{x}|\vec{y})}{\pi(\vec{x})p(s|\vec{x})q(\vec{y}|\vec{x})}.
\label{MHrat}
\end{equation}
Here $\pi(\vec{x})$ is the prior on the parameters and $p(s|\vec{x})$ is the likelihood.  This jump is then accepted with probability $\alpha = {\rm min}(1,H)$, otherwise the chain stays at $\vec{x}$.   The efficiency of any MHMC search is highly dependent on the proposal distributions used.  In previous works, to make small jumps in the parameter space, the most efficient proposal distribution  is a multi-variate Gaussian distribution with jumps that use a product of normal distributions in each eigendirection of the FIM, $\Gamma_{ij}$.  Here we use Latin indices to indicate that this is the FIM on the 5-$D$ subspace of intrinsic parameters.  The standard deviation of the jump in each eigendirection is given by $\sigma_{i} = 1/\sqrt{D\lambda_{i}}$, where $D$ is the dimensionality of the search space (5 in this case) and $\lambda_i$ is the corresponding eigenvalue (the factor of $1/\sqrt{D}$ ensures an average net jump of $\sim 1 \sigma$).   Other similar proposal distributions may also be used to make medium and large jumps.  In addition, due to the response of the LISA detector, it is necessary to include a proposal that jumps to the antipodal sky position, i.e. $\theta\rightarrow\pi-\theta, \phi\rightarrow\phi\pm\pi$.

One of the fastest ways to improve convergence is to ensure that one is using adequate proposal distributions.  However, it has also been shown that by ignoring the Markovian property, the convergence speed of the algorithm can be greatly increased.  The primary modification that increased the convergence speed was the introduction of a number of different annealing schemes.  The first called frequency annealing~\cite{cornishporter1} initially uses a short duration, low frequency waveform ``snippet'', and then gradually increases the frequency band of the template.  The main advantage of this particular annealing scheme is that it avoids a very difficult problem in the detection of SMBHBs.  For bright SMBHBs, most of the SNR comes in the last few days before plunge as the binary enters the highly relativistic regime.  The problem in parameter estimation is that the frequency and phase are increasing very quickly during this phase, and in a high dimensional parameter space it can be very difficult to match.  The frequency annealing avoids this problem by first fitting the early waveform where the frequency evolution is slow.  As the template lengthens and approaches the highly relativistic regime, the chain is then close enough in parameter space to acceptably lock onto the true solution.  In this work we do not use frequency annealing.  However, we do employ the other annealing schemes that we now describe.

\paragraph{Simulated and thermostated annealing}
Simulated annealing works by treating the likelihood like a partition function with an inverse heat.  In the definition of the likelihood, Eq.~(\ref{eqn:likelihood}) we can replace the factor of 1/2 in the exponent with a parameter $\beta$, such that 
\begin{equation}
\beta = \left\{ \begin{array}{ll} \frac{1}{2}10^{-\xi\left(1-\frac{i}{T_{c}}\right)} & 0\leq i\leq T_{c} \\ \\ \frac{1}{2} & i > T_{c}  \end{array}\right.,
\end{equation}
where $\xi$ is the heat-index defining the initial heat, $i$ is the number of the step in the chain and $T_{c}$ is the cooling schedule.  This heats the likelihood surface making it easier to move in an uphill direction.  One of the problems with simulated annealing is knowing, a priori, exactly what the optimal starting temperature should be, especially as each optimal initial heat is source dependent.  If the initial temperature is too hot, we waste too many computer cycles jumping randomly over the likelihood surface.  If it is too low, we quickly converge to a secondary solution and stay there.

To circumvent this problem, the concept of thermostated annealing was introduced~\cite{cornishporter1}.  In this case the heat injected into the likelihood surface becomes a function of the shape of the surface.  The thermostated heat is defined by 
\begin{equation}
\delta = \left\{ \begin{array}{ll} 1.0 & 0\leq SNR\leq \chi \\ \\ \left(\frac{SNR}{\chi}\right)^{2} & SNR > \chi  \end{array}\right. ,
\end{equation}
which ensures that once we reach an SNR greater than $\chi$, the effective SNR never exceeds this value.

\section{Going beyond MHMC}\label{sec:gbmhmc} 
In this section we describe how we can go beyond the MHMC algorithms by constructing a hybrid evolutionary algorithm (HEA).  This involves using concepts such as Nested Sampling, a variant of the MHMC and aspects of evolutionary computation.  We will treat each modification in turn.   

\subsection{Nested sampling}
Nested Sampling (NS)~\cite{Skilling04} was developed as a method for model comparison in Bayesian statistics.  The main advantage of the method was the ability to directly calculate the evidence, something which is notoriously difficult for other algorithms, such as MCMC, to do. The NS algorithm works by randomly populating the parameter space with a number of live points chosen from some prior $\pi(\theta)$, where $\theta$ are the parameters of the system.  The live points are then sorted in order from the lowest to highest likelihood points.  At each iteration $i$, the idea is to find a new point $\vec{x}_{i+1}\in\pi(\theta_i)$ such that $\pi(\theta_{i+1}) \geq \pi(\theta_i)$ and ${\mathcal L}_{i+1}(\theta) > {\mathcal L}_{\rm min}(\theta)$.    Once such a point has been found, the lowest likelihood point in the set is deleted and replaced with the new point.  As a consequence, the entire cluster of points moves to higher likelihood by climbing through nested contours of likelihood, while keeping the number of live points constant.  The new point can be found by uniform, random sampling of the prior.  However, the main difficulty in Nested Sampling is the need to find a higher likelihood point via this random prior sampling.   Depending on the distribution of live points and the shape of the likelihood surface, it can be time consuming to find a new point by random sampling. A more efficient way of sampling higher likelihood points from the prior is known --- we decompose the distribution of live points as a superposition of overlapping ellipsoids and sample from within one of these, chosen at random, and with appropriate weight, from the set. This is the approach employed in the MultiNest algorithm~\cite{multinest}.

\subsection{Metropolis-Hastings nested sampling}
\label{sec:MHNS}
One way to find a higher likelihood point that avoids the need to randomly sample from the prior is to use a MHMC move within the algorithm. At step $i$, when the lowest likelihood in the cluster is ${\cal L}_i$, a point (not necessarily the point of lowest likelihood), with parameters $\Theta$, is chosen at random, and then a short (20-30) iteration Markov chain is used to move that organism, ending at a point with parameters $\Theta'$. This final point is compared to the lowest likelihood point in the cluster (not the original point) and the lowest likelihood is replaced by the point $\Theta'$ with probability $\alpha={\rm min}(1,H)$, where
\begin{equation}
H =  \left\{ \begin{array}{ll} 1 &{\mathcal L}(\Theta')>{\mathcal L}_{i}\,\, \mbox{and}\,\, \pi(\Theta')>\pi(\Theta) \\ \\ \pi(\Theta')/\pi(\Theta) & {\mathcal L}(\Theta')>{\mathcal L}_{i}\,\, \mbox{and}\,\, \pi(\Theta')\leq\pi(\Theta) \\ \\ 0 & \mbox{otherwise}  \end{array}\right. ,
\end{equation}
It is desirable to use a multiple-step MHMC to ensure that the $\Theta'$ is uncorrelated with the initial random point $\Theta$. The new point is included if it has improved the fitness of the cluster and reduced the total prior volume of the point set (i.e., has higher values of the priors).  The new point may also be accepted if it only improves the likelihood.

\subsection{Clustering the organisms}
\label{sec:clus}
One of the key aims in developing these new algorithms was to tackle likelihood surfaces that are multi-modal. An important element in this is to separate out different modes of the solution as they are identified, in order to avoid the whole population migrating to the same peak of the likelihood surface. Mode separation is achieved by clustering the live points periodically into individual groups which are then evolved separately. After a pre-defined number of steps, the clustering is re-evaluated in case modes have merged together or separated further. The clustering is achieved via an algorithm based on the `k-means' and `x-means' techniques~\cite{pelleg}.

In the `k-means' algorithm, the number of modes, $k$, that the set of points will be partitioned into is specified in advance. The point division is then achieved by (i) choosing the first $k$ centres at random; (ii) assigning each point to the centre to which it is closest; (iii) computing the centroid of the points within each cluster and updating the cluster centres to these values; (iv) iterating (ii) and (iii) until a steady state is reached, or a pre-specified number of iterations have been attempted.

The `x-means' algorithm~\cite{pelleg} aims to determine the ``best-fit'' number of clusters that characterizes the live point set. At each step, the k-means algorithm is applied to the point set and the goodness-of-fit of each k-mode decomposition is assessed using a suitable criterion. In our implementation we use the Bayesian Information Criterion (BIC), which is 
\begin{equation}
BIC(M) = \hat{l}_j(D) - \frac{p_j}{2} \cdot \log R
\end{equation}
where $\hat{l}_j(D)$ is the log-likelihood of the data according to model $D$, evaluated at the maximum likelihood point, $R$ is the number of data points and $p_j$ is the number of parameters in the model, $M$. If we represent all the clusters as identical spherical Gaussians, with variance $\hat{\sigma}^2$, the maximum log-likelihood for a cluster containing $R_n$ points is
\begin{equation}
\fl\hat{l}(D_n) = -\frac{R_n}{2}\log(2\pi) - \frac{R_n \cdot N}{2} \log(\hat{\sigma}^2) - \frac{N(R_n-k)}{2} + R_n\log R_n-R_n\log R
\label{biclik}
\end{equation}
where $N$ is the dimensionality of the parameter space and $k$ is the number of means used at this clustering stage. The total log-likelihood is given by the sum of the likelihoods for the $k$ clusters. The number of free parameters, $p_j$, is the sum of $k-1$ class probabilities, $N\cdot k$ centroid coordinates and one variance estimate~\cite{pelleg}.

The user specifies a minimum and a maximum value of $k$ to try. The algorithm is then (i) set $k=k_{\rm min}$; (ii) perform a k-means division and record the BIC score; (iii) for each cluster point set, perform a `2-means' division  of that set; if the BIC of the 2-mode model is higher than the BIC of the 1-mode model, accept the division, otherwise reject it; (iv) evaluate the BIC of the whole resulting set of $k$ clusters and record it; (v) if $k<k_{\rm max}$, set $k=k+1$ and repeat steps (ii)--(v), else stop; (vi) compare BIC scores of all recorded k-means divisions and accept the value of $k$ with the highest BIC.

The likelihood given by Equation~(\ref{biclik}) assumes clusters are identical and spherical, but we have also tried a modified BIC in which each cluster is modeled as an independent ellipsoid. This increases the number of model parameters required to describe each cluster. This did not improve the performance of the algorithm, which is why we continued to employ the spherical version of the BIC. The other refinement we considered was in the choice of distance measure. The easiest distance measure to use when assigning points to clusters and determining the centroid is the Euclidean distance measure on a unit hypercube, i.e., we rescale each coordinate so that they range from $0$ to $1$ over the prior. The natural measure on the waveform parameter space is the Fisher Matrix derived distance, as this describes the expected shape of likelihood peaks, and therefore the shape that we expect clusters to approach at late times. We have implemented a Fisher Matrix version of the clustering algorithm, but it did not show sufficient improved performance to warrant the additional computational overhead. However, the Fisher Matrix does inform us as to which parameters have significant influence on the likelihood, and which parameters have a more minor influence. This allows us to construct a suitable rescaled set of coordinates (not a unit hypercube) for which the likely size of the likelihood peaks in each parameter direction would be comparable. We found this was the better approach to clustering, since it tends to make clusters as spherical as possible, and was the version of the clustering finally employed.

\section{The Hybrid Evolutionary Algorithm}\label{sec:hea} 
Evolutionary algorithms (EAs) have been used extensively in other fields.  These algorithms use aspects from evolutionary biology and natural selection as search techniques in high dimensional spaces.  The most common criteria introduced are concepts like fitness, altruism, selection, leadership, birth, death etc.  A form of evolutionary algorithm, the genetic algorithm, has already been used in gravitational wave data analysis~\cite{ccr} to search for galactic binary systems.  Within the framework of EAs, the candidate solutions play the role of organisms in an environment.  The main advantage of the EA is that it makes no a priori assumptions about the underlying landscape.  These algorithms also have the advantage that they have many organisms travelling over the likelihood surface simultaneously, whereas a standard MCMC algorithm relies on a single chain.  The population evolves according to the principle that each successive generation must be fitter than the last.  The general aim, at each iteration, is to find the weakest member of the population and replace it with a fitter organism.  As we are using a population of organisms, there is a reduced chance that the algorithm will get stuck on a local maximum.   One aspect that we do not use in our EA is mutation.  In genetic algorithms, mutation plays an important role.  This is achieved by representing the candidate solutions in a binary string and changing one of the bits.  However, from evolutionary biology, it is now clear that mutation plays more of a negative than positive role in evolution.  Most mutations result in directing that particular species towards extinction.  Ironically, the main disadvantage of an EA comes from its main advantage.  As we said, an EA makes no prior assumptions about its underlying landscape.  But because of this, it has no concept of the difference between a local and global maximum, nor does it have any natural stopping criterion.  In the following subsections, we describe some of the features of our hybrid evolutionary algorithm (HEA).

\subsection{Fitness criteria for initial selection}
In this study we decided to search a data set with two sources.  Of these sources, one is coalescing within the observation time, and the other coalescing very shortly after the period of observation has finished.  As we knew a priori that the sources had these particular characteristics, we chose our initial set of organisms such that half would search for the coalescing source, and the other half would search for the non-coalescing source.  This places each of the organisms in one of two sub-regions in $t_c$ space.

Our first fitness condition was that, to be accepted, an organism must have a log likelihood corresponding to at least a certain SNR, which we took to be 20 for non-coalescing organisms, or a SNR of 30 or greater for coalescing organisms.  In principle, the choice of initial fitness is down to the user.  We can choose the two fitness criteria to be different from one other from the knowledge that, in general, non-coalescing sources are dimmer.  We should mention that even though we had a priori knowledge of the solutions, our fitness criteria were still set low relative to the SNR of the global solution.  

To begin the search, in each sub-region, we choose the first two points at random, drawing from the prior until the first fitness criterion is satisfied by both organisms.  We then continue to choose points at random, but after the second point, each new accepted organism must satisfy a second fitness condition in addition to the first.  We require that each new organism has a log likelihood which is greater than the mean log likelihood in the sub-region, i.e.  ${\mathcal L}_{n} > {\mathcal L}_{\rm clust}^{\rm mean}$ for $n > 2$.  This ensures that each new organism is fitter than the mean fitness of the local population.  Once the desired number of organisms have been generated, we allow them the chance to improve their fitness by exploring their locality.  For this we move each organism for 300 iterations of an uphill climber MCMC, i.e., jumps are only accepted if the new likelihood is greater than the previous likelihood.  This aims to put each organism onto the nearest peak of the likelihood surface at the start of the search.  The improvement can be small for some live points, but is large for others.   While this initial phase is the most time consuming part of the search, we have found that it is important for the subsequent rapid convergence of the algorithm.  Once this investigative phase is complete, we perform the first partition of the live points.

\subsection{Cluster evolution}
As described in Section~\ref{sec:clus}, we can specify the range of the number of clusters into which the clustering algorithm attempts to partition the live point set. For this particular search we told the algorithm that there were between 2 and 8 possible solutions.  This was equivalent to telling the algorithm that there were between 1 and 4 possible sources in the data, assuming that the algorithm would only find the true solution and the antipodeal sky solution.  If there were less than four sources, it then allowed the algorithm to find other secondary maxima in the likelihood.

\begin{figure}[t]
\begin{center}
\includegraphics[width=3.5in, height=2.75in]{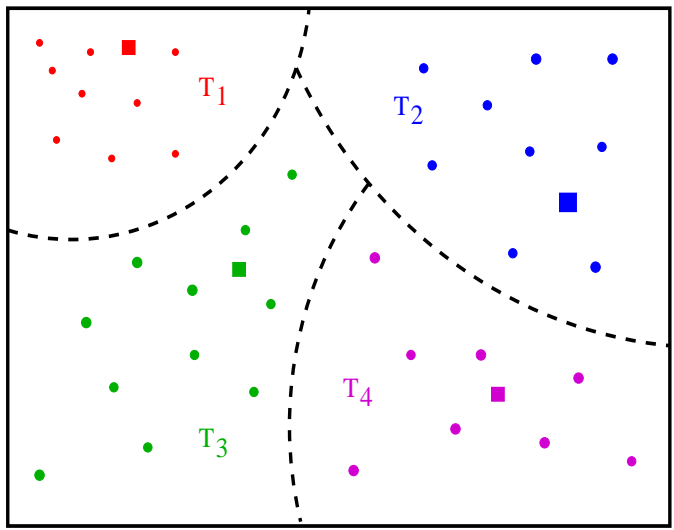}
\end{center}
\caption{A plot of the different Voronoi regions.  In each region we represent the live organisms by circles, and their associated centroids by squares.  Due to the strength of the different solutions in each Voronoi region, we use a different thermostated heating to encourage the organisms to move at their own pace.}
\label{fig:voronoi}
\end{figure}

One of the important things for the evolution of each cluster is the choice of heat used for simulated annealing in that region of the likelihood surface.  Initially all clusters are given the same heat, as we use a common simulated annealing scheme.  However, once the heat for the simulated annealing drops below a certain value, we change this approach.  In geometrical terms, each cluster occupies a Voronoi region of the parameter space (see Figure~\ref{fig:voronoi}).  As each cluster will eventually be associated with a mode of the likelihood of different brightness, it does not seem logical to subject each cluster to the same heat.  To improve convergence, we allowed each Voronoi region to have a different thermostated heat depending on the SNR of the fittest member of each cluster.  This allows the fitter clusters to move faster through the parameter space.  As all the clusters improve in fitness, the heat equilibrates between the Voronoi regions associated with the same source.

Another advantage of the different heats in the different Voronoi regions is that the clusters evolve independently.  In other words, there is no inter-cluster competition.  All competition between organisms occurs only within a cluster.  At each step, we update the clusters in one of three ways.  The first is a straight-forward update, where we try to update the lowest likelihood organism of each cluster by a random choice of parameters within a certain distance of its current location.  To do this, we construct a hypercube in the parameter space around the current point, where the distance from the current point to the surface of the hypercube is given by 
\begin{equation}
D = \delta \sqrt{h_c  C^{\mu\mu}}
\end{equation} 
where $C^{\mu\mu}$ are the diagonal elements of the inverse FIM, $h_c$ is the heat associated with that particular cluster and $\delta$ is a scaling factor.  The scale factor decreases the volume of the hypercube in a monotonic fashion as we make subsequent trials.  The reason for this scaling is to encourage the lowest likelihood organism to improve its fitness by allowing it to look initially at some distance away, but eventually looking in its nearby environment as well.  For this particular move we allow the organism 100 attempts to become fitter, assigning $\delta$ values between 1 and 1/100.

The second option for a cluster to move is to choose an organism at random within the cluster and update it using the MHMC technique described in section~\ref{sec:MHNS}.  If we cannot improve the likelihood this way, we try to update the lowest likelihood organism in a similar manner.  In both cases we allow the organism 50 attempts with the directed proposal distributions to improve their fitness.

The third option is to try and sample a new point uniformly from within an iso-likelihood contour passing through the lowest likelihood live point of the set. To achieve this, we find the best fit ellipsoid to the current live point set, and assume that this represents the shape of the likelihood contours. We then draw uniformly from within the ellipsoid through the lowest likelihood point until a higher likelihood point is found.

\subsection{Immigration/Emigration.}
While there is no inter-cluster competition, we re-cluster the organisms after a predetermined number of iterations (20 in this analysis), which allows immigration/emigration between clusters.  This also allows cluster numbers to change over the course of the run.  In general we find that each cluster has an optimal population size.  If the cluster is too small or too big, it does not evolve as quickly as other clusters.  One aspect that we may explore in the future is to examine the global fitness of a cluster, as well as the fitness of the individual organisms.  In cases where we find the rate of increase of the overall fitness of the cluster beginning to slow, we could induce forced immigration/emigration.

\subsection{Elitism}
The concept of elitism has been very important in evolutionary biology and is also becoming important in evolutionary algorithms.  In this algorithm we single out the fittest member of each cluster and allow them the opportunity to improve their current fitness level.  This accelerates the convergence of the algorithm over a straight-forward Nested Sampling search.  In that case, the fittest member has to be selected at random, or another organism may become the fittest.  In our algorithm, being the fittest is rewarded with special treatment over all others.  In this case the fittest member of each cluster is never replaced and in fact, guides the direction of the entire cluster.  We should point out that we do allow competition within each cluster, so that while the fittest organism is given special treatment,  other organisms can become the fittest member of the cluster.  This competition prevents a runaway scenario where a cluster evolves too quickly and gets stuck on a secondary solution.

\subsection{Altruism and crossover}
We also attempt to improve clusters with the ideas of altruism and crossover.  Altruism is normally defined as a selfless concern for the welfare of others, whereas crossover is defined as the creation of a new organism using information from existing organisms.  In this context we use altruism as follows : we assign a particular fitness criterion to the fittest members of any cluster associated with a particular solution.  If one cluster is not performing very well, the fittest member of the fittest cluster is moved to the under-performing cluster.  The crossover then comes in as follows: The now fittest member breeds with the second fittest member (this is done very simply by exchanging combinations of chirpmass, reduced mass, time to coalescence, or a mixture of all three).  The new organism should now be fitter than the weakest member of the cluster.  If this is the case, the weakest member is replaced.  If not, the fittest member is reabsorbed into a better performing population during the next round of clustering. 

\subsection{Reduction of the parameter space volume}
One of the most crucial aspects of all evolutionary algorithms is the ability to reduce the volume of the priors in an efficient manner.  At the beginning of the algorithm we divide the organisms into coalescing and non-coalescing solutions, thus splitting the total volume of the parameter space.  Once the centroids of each cluster start to converge to local solutions, we use their fitness information to decide on whether or not they should guide the other clusters.  For this we proceed as follows: as each cluster moves through the parameter space, we examine the fitness of the cluster members.  The fitness of the worst and best organisms define the prior volumes for the source parameters.  We also examine the values of the times of coalescence to see if we can associate each cluster with the same source.  We call clusters part of the same family if the difference between the fittest members of each cluster is less than $10^{-3}$ years.  This corresponds to about 8.8 hours (while always possible, it seems unlikely that two sources would be coalescing in the same $\sim9$ hour period).  If we can associate two separate clusters as being part of the same family of solutions, we ensure that the priors for the best performing cluster are included in the priors for the other family members as well.  This prevents a cluster confining itself to a distant solution without the possibility of improving.

\section{Results}\label{sec:heares} 
\begin{figure}[t]
\begin{center}
\includegraphics[width=\textwidth, keepaspectratio=true]{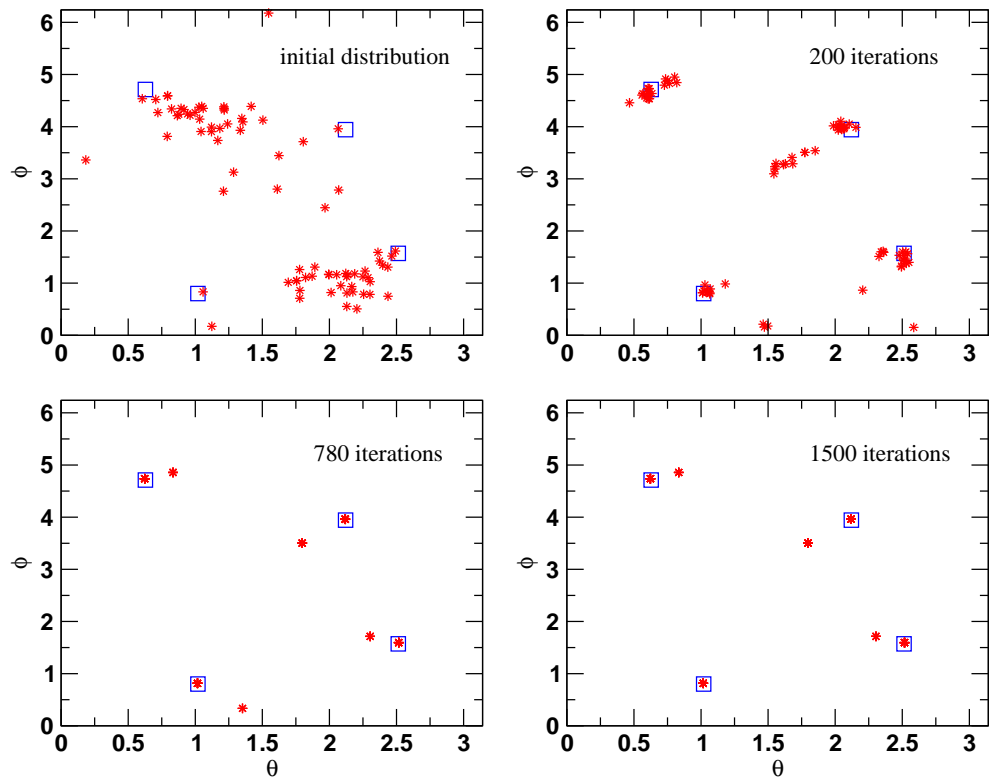}
\end{center}
\caption{A plot of the sky search at four different times.  In each cell, the square represent the true solutions, while the stars represent the organisms.  We plot (going from top-left to bottom-right) the initial distribution after the initial selection and uphill climber improvement phase, and then at 200, 780 and 1500 iterations.  We see that not only does the algorithm find the two primary sky solutions (i.e., real and antipodal) for both sources, but also a bunch of secondary solutions at almost 90 degrees to the primaries.  The range of the axes is the same in all plots.}
\label{fig:skysearch}
\end{figure}

\begin{figure}[t]
\begin{center}
\includegraphics[width=\textwidth, keepaspectratio=true]{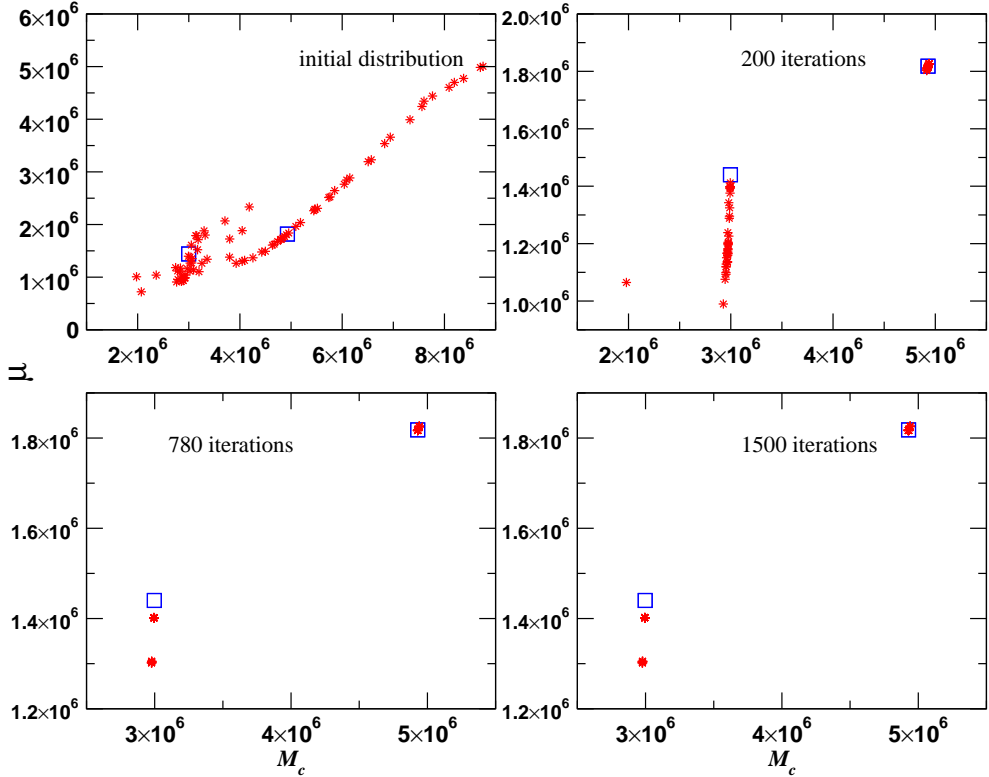}
\end{center}
\caption{A plot of the search over the chirp and reduced masses (in units of solar masses) for the same four time stamps.  Initially there is a wide distribution of points. However, we can see from the shrinking axes that we very quickly converge on the true solutions.  While we do well with chirp-mass, we are still a little off on reduced-mass. }
\label{fig:masssearch}
\end{figure}

\begin{figure}[t]
\begin{center}
\includegraphics[width=\textwidth, keepaspectratio=true]{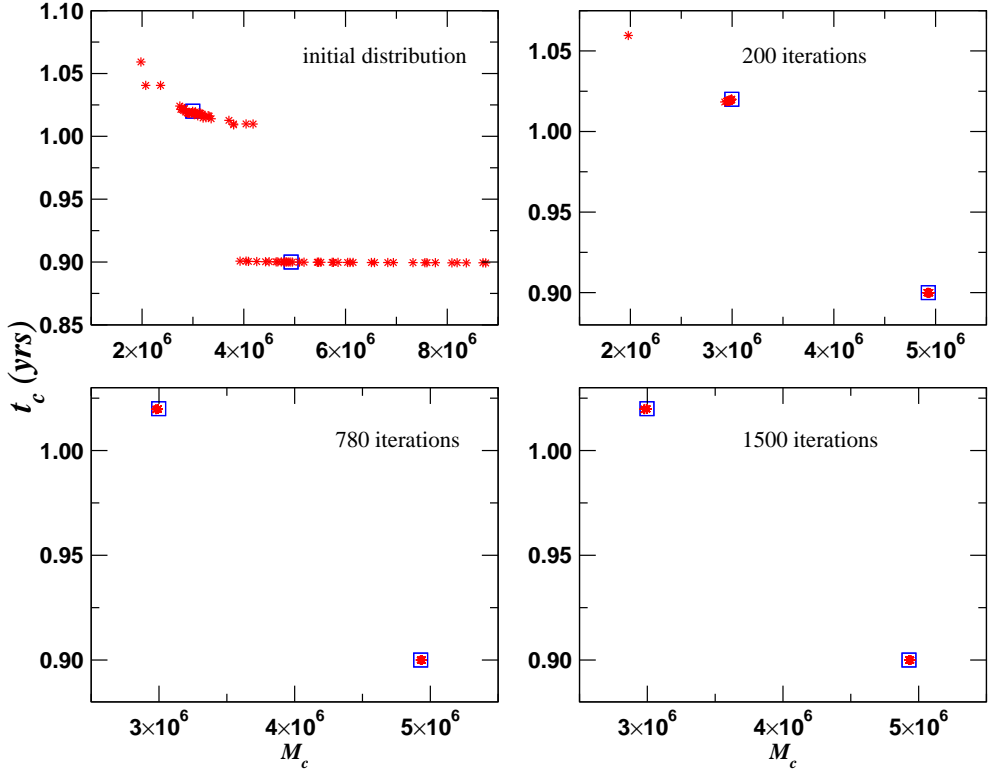}
\end{center}
\caption{A plot of the search over chirp-mass and time to coalescence.  We can see that the initial fitness criteria spread the points over a wide range of chirp-masses for the bright source, but along a tight prior for $t_c$.  For the dim source, there is a greater spread in both parameters.  However, by 200 iterations, we have converged on the correct answer for both sources, and we have one extra cluster associated with the dim source.  By 780 iterations, this cluster has also migrated to the primary solution.  Note once again the shrinking axes in each cell.}
\label{fig:masstimesearch}
\end{figure}

We ran the search algorithm for 1500 iterations using 80 live organisms on a data set containing the two sources listed in Table~\ref{tab:params}.  To recap the algorithm, the initial search divides the number of organisms between coalescing and non-coalescing organisms.  The original population of organisms is selected so that each successive organism is fitter than the current mean fitness of the population.  Once the population has been created, the organisms are allowed the opportunity to improve their fitness using an uphill climber proposal.  We then create a number of centroids at random and allow the organisms to be clustered into sub-populations.  Afterwards, the clusters evolve according to the methods outlined in the previous sections.  We point out once again, that all inter-organism competition is confined to individual clusters and there is no inter-cluster competition.  During the first 750 iterations we use a mixture of simulated and thermostated annealing, while for the final 750 iterations we use a slow cool-down until unit heat was reached.  In all cases we allow the temperature in each Voronoi region to evolve independently.  In Figures~\ref{fig:skysearch}-\ref{fig:masstimesearch} we show the evolution of the live point set as it searches for the two sources.  In each plot we have four cells.  Going from left to right, top to bottom, we plot the parameters at the following points : (a) after the first clustering has taken place, (b) after 200 iterations, (c) after 780 iterations and (d) the final positions in the parameter space.

In Figure~\ref{fig:skysearch} we track the evolution of the organisms in the $(\theta,\phi)$ subspace.  We can see that even after the initial selection criteria has been satisfied and the first clustering has taken place, the organisms are spread throughout the parameter space, although some of the fitter organisms are already close to the main modes of the solution.  After two hundred iterations, most of the organisms have started to cluster around the two modes (i.e., the real and antipodal sky locations) of each source, or one of several secondary modes.  By 780 iterations, the original population has fragmented into eight individual clusters.  Of these, four match the expected modes of the solution.  The other four clusters have matched secondary solutions.  What is interesting about the solutions at this point is that the sky modes appear in reflective pairs, with the secondary pair rotated to almost $\pi/2$ from the primary modes. This also happens to be approximately the sky-location of the other source in each case, i.e., there is a secondary mode of the coalescing source at roughly the sky position of the non-coalescing source and vice-versa. This is an interesting feature of the multi-source likelihood surface that warrants further investigation. By the end of the run, the primary modes have refined their solutions, and one of the secondary modes of source 1 has migrated to one of the primary modes.  Figure~\ref{fig:masssearch} shows the evolution in the $(M_c,\mu)$ subspace.  In this case we see that, even with a quite competitive initial fitness criterion, there is still a wide spread in the possible masses of the organisms, with chirp masses ranging from $2\leq M_c\times10^6/M_{\odot}\leq9$ and reduced mass ranging from $1\leq \mu\times10^6/M_{\odot}\leq5$.  After 200 iterations the volume of the parameter space has shrunk dramatically.  The organisms have already converged on the true solutions for the bright source.  It is clear that while the organisms have also converged to the correct chirp mass, there is quite a spread in reduced mass for the dimmer source.  By 780 and 1500 iterations, the organisms have begun to converge and in fact, only carry out a small refinement during this period.  In Figure~\ref{fig:masstimesearch}, we plot the evolution in the $(M_c,t_c)$ subspace.  In this case we can see that the initial fitness criteria has had an interesting effect.  For the bright source, while there is a large spread in chirp mass, all the organisms have approximately the correct coalescence time. This is presumably because a significant fraction of the SNR accumulates near coalescence. For the dimmer source, as the coalescence is not seen, there is a greater spread in both parameters.  After 200 iterations a number of the clusters have converged to the correct solutions, although one cluster associated with the dim source has moved toward the edge of the prior, with a low mass value and a coalescence time of approximately 1.05 years.  However, by 780 iterations, all the organisms have converged to the correct solutions in this subspace, and the solutions are then further refined until the end of the run.

In order to give a quantitive assessment of the algorithm, in Table~\ref{tab:parerrs} we quote the predicted one sigma errors from the Fisher matrix.  We also quote the errors for the antipodal sky solutions having taken care of the appropriate rotations for the inclination and polarization angles.  Note that as this search is based on the F-Statistic, we only quote errors for the five parameters for which we search.  In Table~\ref{tab:eaerrs} we give the errors in the recovered parameters as multiples of the Fisher Matrix error estimate, i.e $|(\lambda_T - \lambda_{\rm MAP})/\sigma_{\rm FIM}|$, where $T$ denotes true value, MAP denotes maximum a posteriori value and $\sigma_{\rm FIM}$ is the Fisher matrix one sigma error prediction.  We can see that the algorithm did extremely well on the bright source, with all recovered parameters within $3\sigma$ of the true values.  The performance for the dim source, while not as impressive, was still good, with all recovered parameters within $5\sigma$ of the true answer.  While it is clear that there is still some tuning to do, to our knowledge, this is the first time that a simultaneous detection of SMBHBs has been attempted for LISA.   The important conclusion is that not only did we find the two sources, but we also found all the modes of each source.  In terms of run time, the algorithm took about ten hours to run on a MacBook Pro with a 2.6 GHz dual-core processor and 4 Gb of memory.  Of these ten hours, about four hours are spent refining the initial population. 

\begin{table}
\caption{\label{tab:parerrs}Fisher matrix predictions for parameter errors.  Note that we also quote the theoretical antipodal sky solution error estimates (denoted by A).} 

\begin{indented}
\lineup
\item[]\begin{tabular}{@{}*{6}{l}}
\br                              
source&$\sigma_{M_c}/M_{\odot}$&$\sigma_{\mu}/M_{\odot}$&$\sigma_{t_c}/$yrs&$\sigma_{\theta}/$rads&$\sigma_{\phi}/$rads\cr 
\mr
1&$1.289\times10^3$ &$4.719\times10^3$&$1.209\times10^{-5}$&$6.315\times10^{-3}$&$1.159\times10^{-2}$ \cr
1A&$1.198\times10^3$&$4.405\times10^3$&$1.128\times10^{-5}$&$9.683\times10^{-3}$&$8.529\times10^{-3}$  \cr 
2&$6.986\times10^2$ &$8.642\times10^3$&$3.292\times10^{-5}$&$6.283\times10^{-3}$ &$7.854\times10^{-3}$\cr
2A&$7.025\times10^2$&$8.683\times10^3$&$3.301\times10^{-5}$&$6.446\times10^{-3}$&$7.631\times10^{-3}$  \cr 
\br
\end{tabular}
\end{indented}
\end{table}

\begin{table}
\caption{\label{tab:eaerrs} Errors in parameter recovery for the two test sources, quoted as $|(\lambda_{T}-\lambda_{\rm MAP})/\sigma_{\rm FIM}|$, where $T$ denotes the true value, MAP denotes the maximum a posteriori value and $\sigma_{\rm FIM}$ is the one sigma estimation from the Fisher matrix.  Note that while we have found multiple modes of the solution, we are only quoting errors for the true and antipodal sky solutions.} 

\begin{indented}
\lineup
\item[]\begin{tabular}{@{}*{6}{l}}
\br                              
$$&$ M_c$&$\mu$&$tc$&$\theta$&$\phi$\cr 
\mr
1&$0.2792$ &$0.1993$&$0.0579$&$0.8477$ &2.1338\cr
1A&$0.4637$&$0.2813$&$1.0106$&$0.3733$&2.8196  \cr 
2&$3.6418$ &$4.3916$&$4.3870$&$0.2015$ &2.4197\cr
2A&$3.7056$&$4.3759$&$3.2116$&$0.6865$&1.7567  \cr 
\br
\end{tabular}
\end{indented}
\end{table}

\section{Conclusion}\label{sec:discuss} 
We have developed a Hybrid Evolutionary Algorithm and applied it to the problem of simultaneous detection of two SMBHBs in LISA data.  The algorithm uses aspects from Nested Sampling, Metropolis-Hastings theory and evolutionary computation to evolve a population of organisms through the parameter space.   After an initial selection based on a particular fitness criterion, the organisms are clustered into sub-populations. We then evolve these clusters according to a number of different criteria.  While we do not allow inter-cluster competition, we do permit emigration/immigration between clusters and all inter-organism competition takes place on a local level. Each cluster occupies a Voronoi space, and in each space we use a combination of simulated and thermostated annealing.  By allowing each Voronoi region to have its own heat, we find that the clusters evolve more rapidly.

Using this hybrid algorithm, we were able to detect the two injected SMBHBs simultaneously.  As well as detecting the true sky solution mode, we were also able to detect the antipodal sky modes.  Furthermore, we also detected secondary mode solutions for each of the sources.  In both cases the maximum a posteriori parameter values were within $5\sigma$ of the injected values.  At present the algorithm uses 80 organisms, and the entire run takes about 10 hours on a MacBook Pro laptop.  The code in its current state uses a number of aspects that can be trimmed in the future, which should make it possible to reduce the run time by at least $50\%$.

It is clear that this algorithm has a lot of potential and is worth developing further in the context of gravitational wave astronomy. It is not immediately clear how the algorithm's performance will scale when used to search for other LISA source types, such as spinning black holes or EMRIs, where the likelihood surface contains many more secondaries. We plan to explore the application of this technique to these other cases in the future. 

The HEA  was originally based on the Nested Sampling algorithm described in~\cite{feroz08}, but that work has now been superceded by the improved MultiNest algorithm~\cite{multinest}. The MultiNest algorithm is entirely model-independent in the sense that the problem to which it is applied enters specifically only in the computation of the likelihood and not in the techniques used to update the live point set. It utilizes an efficient way to sample higher likelihood points from the prior by decomposing the distribution of live points as a superposition of overlapping ellipsoids. One of these ellipsoids is chosen at random, with appropriate weight, from the set and a new point is sampled uniformly from within it. The algorithm has proven to be very effective in many different applications~\cite{multinest}. We are also in the process of exploring the application of this algorithm to gravitational wave data analysis problems, and these results will be reported elsewhere.
\\

\ack We thank Farhan Feroz and Mike Hobson for first pointing us to the Nested Sampling algorithm, and for very useful discussions and collaboration on this project. JG's work is supported by a Royal Society University Research Fellowship.

\end{document}